\documentclass{mnras}

\usepackage{graphicx}
\usepackage{footnote}
\usepackage[font=small]{caption}
\usepackage{indentfirst}
\usepackage{tabu}
\usepackage[percent]{overpic}
\usepackage{transparent}
\usepackage{ragged2e}
\usepackage[utf8]{inputenc}

\let\bibhang\relax
\usepackage{natbib}
\usepackage{booktabs}
\usepackage{fixltx2e}

\usepackage{url}
\usepackage{hyperref}

\usepackage{amssymb}
\usepackage{txfonts}
\usepackage[a]{esvect}
\usepackage{tikz}

\newcommand{\degree}[0]{$^{\circ}$}

\newcommand{\cbra}[1]{\left( #1 \right)}      % put the argument between parentheses (curve brackets)
\let\baraccent=\=
\renewcommand{\=}[1]{\stackrel{#1}{=}} % for putting numbers above =
\let\arrowaccent=\>
\renewcommand{\>}[1]{\stackrel{#1}{\Rightarrow}} % for putting numbers above =>

\makeatletter
\newcommand{\rmnum}[1]{{\footnotesize{\expandafter\@slowromancap\romannumeral #1@}}}
\newcommand{\Rmnum}[1]{{\expandafter\@slowromancap\romannumeral #1@}}
\makeatother

\everymath={\rm}
\newcommand{\tfm}[1]{$^{#1}$}
\newcommand{\tablefoottext}[2]{$^{(#1)}$ #2}

\hypersetup{
   colorlinks=true,       % false: boxed links; true: colored links
   linkcolor=black,       % color of internal links (change box color with linkbordercolor)
   citecolor=blue,        % color of links to bibliography
   urlcolor=black         % color of external links
}

\begin{document}

\title[Galaxy interaction in IRAS\,17020+4544]{Evidence of galaxy interaction in the Narrow-line Seyfert 1 galaxy IRAS\,17020+4544 seen by NOEMA}

\author[Salom\'e et al.]{
   % List of authors
   Q. Salom\'e$^{1}$,
   A. L. Longinotti$^{1,2,3}$,
   Y. Krongold$^{3}$,
   C. Feruglio$^{4,5}$,
   V. Chavushyan$^{1}$,
   \newauthor
   O. Vega$^{1}$,
   S. García-Burillo$^{6}$,
   A. Fuente$^{6}$,
   A. Olguín-Iglesias$^{7}$,
   V. M. Patiño-Álvarez$^{1,8}$,
   \newauthor
   I. Puerari$^{1}$,
   and A. Robleto-Or\'us$^{9}$
\\
   % List of institutions
   $^{1}$ Instituto Nacional de Astrof\'isica, \'Optica y Electr\'onica, Luis E. Erro 1, Tonantzintla 72840, Puebla , Mexico \\ email: qsalome@inaoep.mx \\
   $^{2}$ CONACyT-INAOE \\
   $^{3}$ Instituto de Astronom\'ia, Universidad Nacional Aut\'onoma de M\'exico, Circuito Exterior, Ciudad Universitaria, Ciudad de M\'exico 04510, Mexico \\
   $^{4}$ INAF Osservatorio Astronomico di Trieste, via G. Tiepolo 11, 34143 Trieste, Italy \\
   $^{5}$ IFPU-Institute for Fundamental Physics of the Universe, via Beirut 2, 34014 Trieste, Italy \\
   $^{6}$ Observatorio Astron\'omico Nacional (OAN-IGN)-Observatorio de Madrid, Alfonso XII 3, 28014 Madrid, Spain \\
   $^{7}$ Universidad Aut\'onoma del Estado de Hidalgo, \'Area Acad\'emica de Matem\'aticas y F\'isica, Instituto de Ciencias B\'asicas e Ingenier\'ia, \\ Carretera Pachuca-Tulancingo Km. 4.5, Mineral de la Reforma 42184, Hidalgo, Mexico \\
   $^{8}$ Max-Planck-Institut f\"ur Radioastronomie, Auf dem H\"ugel 69, 53121 Bonn, Germany \\
   $^{9}$ Departamento de Astronom\'ia, Universidad de Guanajuato, Apdo. 144, Guanajuato 36000, Gto., Mexico
}

\date{Accepted 2020 November 13. Received 2020 November 11; in original form 2020 September 25}

\pubyear{2020}

\label{firstpage}
\pagerange{\pageref{firstpage}--\pageref{lastpage}}
\maketitle

\begin{abstract}
   The narrow-line Seyfert 1 galaxy IRAS\,17020+4544 is one of the few sources where both an X-ray ultra-fast outflow and a molecular outflow were observed to be consistent with energy conservation. However, IRAS\,17020+4544 is less massive and has a much more modest active galactic nucleus (AGN) luminosity than the other examples. Using recent CO(1-0) observations with the NOrthern Extended Millimeter Array (NOEMA), we characterised the molecular gas content of the host galaxy for the first time. We found that the molecular gas is distributed into an apparent central disc of $1.1\times 10^9\: M_\odot$, and a northern extension located up to 8 kpc from the centre with a molecular gas mass $M_{H_2}\sim 10^8\: M_\odot$.
The molecular gas mass and the CO dynamics in the northern extension reveal that IRAS\,17020+4544 is not a standard spiral galaxy, instead it is interacting with a dwarf object corresponding to the northern extension. This interaction possibly triggers the high accretion rate onto the super massive black hole.
Within the main galaxy, which hosts the AGN, a simple analytical model predicts that the molecular gas may lie in a ring, with less molecular gas in the nuclear region. Such distribution may be the result of the AGN activity which removes or photodissociates the molecular gas in the nuclear region (AGN feedback).
Finally, we have detected a molecular outflow of mass $M_{H_2}=(0.7-1.2)\times 10^7\: M_\odot$ in projection at the location of the northern galaxy, with a similar velocity to that of the massive outflow reported in previous millimeter data obtained by the Large Millimeter Telescope.
\end{abstract}


\section{Introduction}

   Active galactic nuclei (AGN) are assumed to play a major role in the evolution of galaxies. In particular, AGN feedback is often invoked as a key ingredient in regulating star formation. A viable feedback agent that is often proposed is the powerful outflows which eject gas from the innermost regions out to larger scales (see \citealt{Harrison_2018}, and references within). These winds transport  kinetic energy to the entire galaxy, that finally is converted into mechanical energy that heats the gas. As a consequence, star formation is quenched due to a lack of cold molecular gas reservoir. AGN-driven outflows can be observed in X-ray spectra as ultra-fast outflows (UFO at $v>10^4\: km\,s^{-1}$; \citealt{Tombesi_2012,Gofford_2013,Longinotti_2015}), in the optical band (e.g. \citealt{Harrison_2014}) or in the molecular phase (as probed by mm-observations; \citealt{Cicone_2014,Aalto_2017,Alonso_2019,Audibert_2019,Garcia_2019}).

   To date, there are only few sources where both an X-ray UFO and a molecular outflow were observed and physically related \citep{Feruglio_2015, Tombesi_2015,Veilleux_2017,Longinotti_2018a,Bischetti_2019}. Three of these sources are particularly interesting as the two outflow phases are found to follow an energy-conserving relation (see the review of \citealt{King_2015}): Mrk 231, IRAS\,F11119+3257 and IRAS\,17020+4544.

   Contrary to the former two objects which are ultra-luminous infrared galaxies (ULIRG), the narrow line Seyfert 1 (NLSy1) galaxy IRAS\,17020+4544 (hereafter IRAS17) is less luminous, contains less molecular gas and the central AGN is less powerful ($L_{AGN}=5\times 10^{44}\: erg\,s^{-1}$; \citealt{Longinotti_2018a}).
The central AGN is known to be characterized by a warm absorber seen as X-ray absorption \citep{Leighly_1997,Komossa_1998,Vaughan_1999} and by compact radio emission \citep{Snellen_2004}. Using the B-magnitude, \cite{Wang_2001} derived the mass of the central super massive black hole: $M_{BH}=5.9\times 10^6$.

   A recent rich multi-wavelength campaign for IRAS17 has revealed: (1) the presence of the sub-relativistic X-ray UFO \citep{Longinotti_2015} and of a massive, possibly galaxy-scale molecular outflow \citep{Longinotti_2018a}; (2) a non-thermal apparent milli-arcsec scale jet which seems to coexist with the different phases of the outflow \citep{Giroletti_2017}; (3) multi-component X-ray winds \citep{Sanfrutos_2018} which can be interpreted as an evidence that a cocoon of hot gas is producing large-scale shocks in the interstellar medium (ISM).

\begin{table}
  \centering
  \large
  \caption{\label{table:IRAS17020} General properties of IRAS\,17020+4544}
  \begin{tabular}{lc}
    \hline \hline
    RA (J2000)                 &  $17^h 03^m 30^s.383$\tfm{a}  \\
    Dec (J2000)                &     $+$45:40:47.17\tfm{a}     \\
    z                          &         0.0612\tfm{b}         \\
    $D_L$ (Mpc)                &             274.3             \\
    Scale (kpc/$''$)           &             1.181             \\ \hline
    $M_{BH}$ ($M_\odot$)       &   $5.9\times 10^6$\tfm{\,c}   \\
    $L_{AGN}\: (erg\,s^{-1})$  &   $5\times 10^{44}$\tfm{\,d}  \\
    $L_{IR}\: (L_\odot)$       &  $3.1\times 10^{11}$\tfm{\,e} \\
    $SFR\: (M_\odot\,yr^{-1})$ &           26\tfm{f}           \\ \hline
  \end{tabular} \\
  \justify {\small
  \textbf{Notes.}
    \tablefoottext{a}{Position of the unresolved 8.4 GHz emission \citep{Doi_2007}} \hspace{1em}
    \tablefoottext{b}{Derived from the CO emission}
    \tablefoottext{c}{\cite{Wang_2001}}
    \tablefoottext{d}{\cite{Longinotti_2018a}}
    \tablefoottext{e}{\cite{Lee_2016c}}
    \tablefoottext{f}{\cite{Murphy_2011,Kennicutt_2012}} \\
    The IR luminosity was corrected for the AGN contamination before calculating the SFR.
  }
\end{table}

   Contrary to the central AGN which is widely studied, little is known about its host galaxy. The morphology of the galaxy remains debated. Based on observations from the University of Hawaii 88 inch telescope, \cite{Ohta_2007} concluded that IRAS17 is a barred spiral. On the contrary, using B- and R-bands observations from the Nordic Optical Telescope (NOT), \cite{Olguin_2020} do not report complex morphology.
\cite{Lee_2016c} studied the infrared (IR) spectral energy distribution of a sample of 14 galaxies. They derive an IR luminosity between 8 and $1000\: \mu m$ of $L_{IR}\sim 3.1\times 10^{11}\: L_\odot$ for IRAS17 (LDOG-45 in their article), with a contribution of the AGN of about 44\% of the IR luminosity, which led them to classify it as a DOG (Dust Obscured Galaxy). Table \ref{table:IRAS17020} summarizes general properties of IRAS17.

   In this paper, for the first time we characterise the molecular gas of the galaxy using new NOrthern Extended Millimeter Array (NOEMA) observations of the CO(1-0) line. This paper is organised as follow. The observations are presented in Section \ref{sec:Obs}. In Sections \ref{sec:Gas_content} and \ref{sec:CO_kin}, we analysed the molecular gas content and the CO kinematics, respectively. Finally, our results are discussed in Section \ref{sec:conclusion}.
Throughout this paper, we assume $H_0=70\: km\,s^{-1}\,Mpc^{-1}$, $\Omega_M=0.3$, $\Omega_{vac}=0.7$.

\section{Observations}
\label{sec:Obs}

   \subsection{NOrthern Extended Millimeter Array}

   IRAS17 was observed in CO(1-0) with NOEMA (project W17CR; PI: Pati\~no-\'Alvarez). The observations were carried out in April-May 2018 with 9 antennas in C configuration (baselines from 15 to 290m) during 4.8h on-source. The bright QSO 3C454.3 (9.3 Jy at 108 GHz) was used as RF calibrator for the receiver bandpass calibration, while the QSOs 1726+455 and 1656+482 were used as phase/amplitude calibrators. Finally, the star MWC349 was used as absolute flux calibrator. Absolute flux calibration has an uncertainty of about 10\%.

   The calibration of phase, absolute flux and amplitude was performed using the GILDAS software\footnote{\url{http://www.iram.fr/IRAMFR/GILDAS}}. Then, we imaged and cleaned the data using the GILDAS package {\tt mapping}. We first extracted the continuum in the {\tt uv} plane by averaging the channels which do not contain CO emission.
We have imaged the continuum emission at 3mm and found that it is unresolved at a resolution of $2.2''\times 1.7''$. Using the GILDAS task {\tt uv\_fit}, we have fitted a point source in the {\tt uv} plane. The continuum flux density peaks at an offset of $[-0.3,0.0]$ arcsec, relative to the position of the unresolved 8.4 GHz emission from \cite{Doi_2007}, with a total flux of $1.9\pm 0.3\: mJy$ at 108 GHz. %0.07+0.19

   The observations covered a large bandwidth of 8 GHz at a nominal spectral resolution of 2 MHz. In this paper, we focus on the molecular gas within the galaxy. We therefore select a conservative velocity range $-1000<v<1000\: km\,s^{-1}$ to have all the CO emission. After subtracting the continuum within the {\tt uv} plane, we decreased the spectral resolution to $4\: MHz\sim 11\: km\,s^{-1}$ in order to improve the signal-to-noise without affecting our study of the kinematics. We used the {\tt hogbom} method with a threshold at $2\sigma$ to clean the data, where $\sigma$ is the noise given by the deconvolution $\sim 0.98\: mJy/beam$ for channels of $\sim 11\: km\,s^{-1}$. The final datacube has a resolution of $2.23''\times 1.70''$ ($PA\sim 43$\degree).

   \subsection{Large Millimeter Telescope}

   IRAS17 was previously observed with the Large Millimeter Telescope Alfonso Serrano (LMT), located on the Sierra Negra Volcano in Mexico. The observations were carried out on May-June 2015 with the Redshift Search Receiver instrument (RSR), which covers the frequency range 73-111 GHz with a spectral resolution of 31 MHz. The galaxy was observed for a total of 525 minutes, giving an overall rms of 0.2 mK. For more details on the observations and data reduction, we refer to \cite{Longinotti_2018a}.

   \subsection{Optical spectra}
   \label{sec:z_opt}

%%%%%%%%%%%%%%%%

\begin{figure*}
  \centering
  \includegraphics[width=0.95\linewidth]{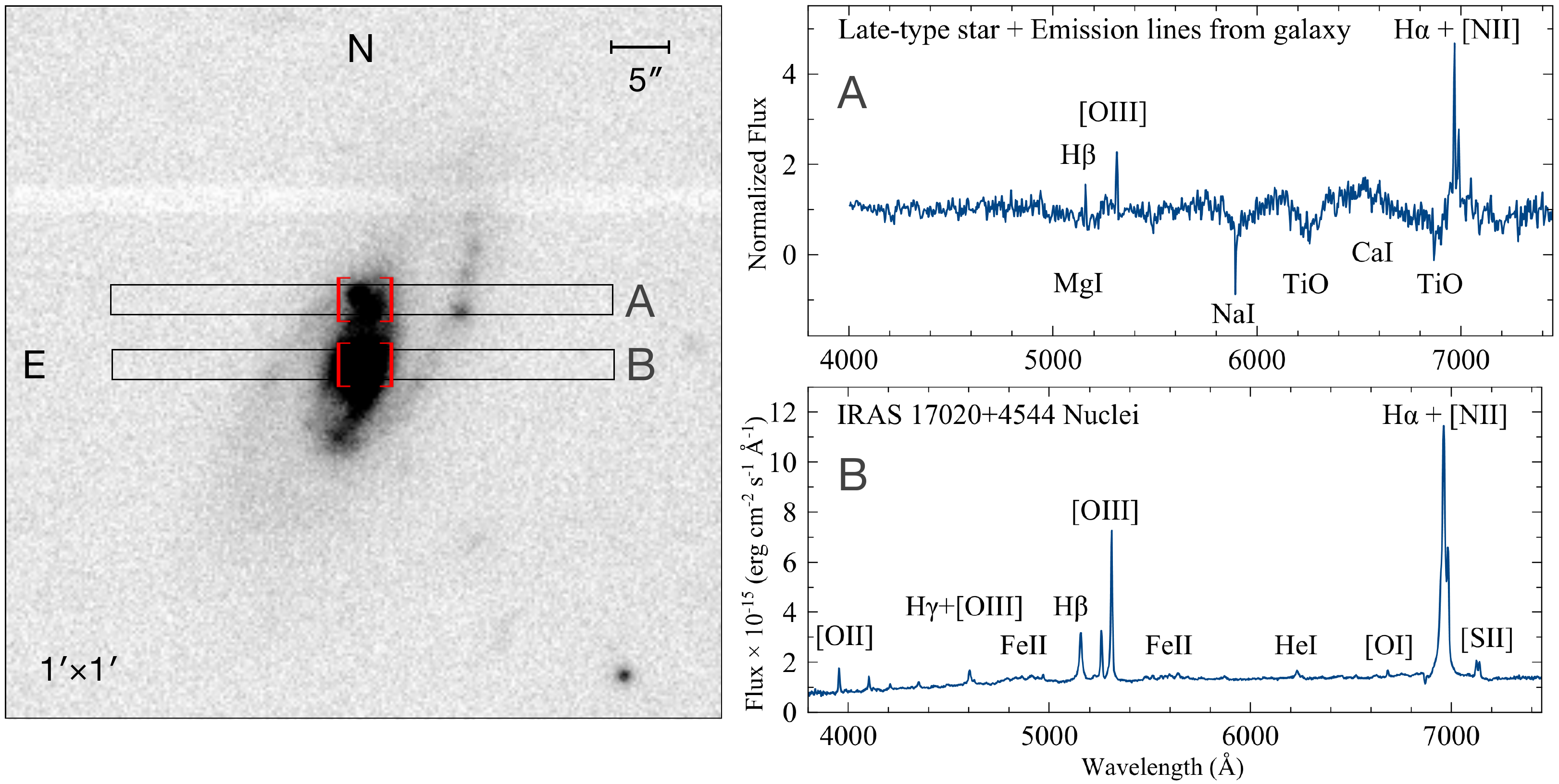}
  \caption{\label{optical} \emph{Left:} B band image taken by the Nordic Optical Telescope \citep{Olguin_2020}. The position of the two slits are shown. The regions from which the spectra were extracted are marked in red.
  \emph{Right:} Optical spectra from the central (B; \emph{bottom}) and northern regions (A; \emph{top}). Spectrum A was not calibrated in flux as observing the spectrophotometric standard was not possible due to bad sky conditions. For that reason the spectrum is shown normalised to the continuum.}
\end{figure*}

%%%%%%%%%%%%%%%%

   The nuclear region of the galaxy IRAS\,17020+4544 and a northern optical-bright spot were observed on 13 March 2015, and 09 May 2016 respectively with the 2.12-m telescope of the San Pedro M\'artir Observatory in Mexico.  It used a Boller \& Chivens spectrograph and a $2048\times 2048$ pixels E2V-4240 CCD, which was tuned approximately to the $3800-8000\: \AA$ range. A 2.5 arcsec slit was used and the position of the slit for each observation is presented in Figure \ref{optical}.
The exposure time is $3\times 1800s$ for the slit centred on the nuclear region, and 1200s for the slit in the north. The spectral resolution is $8\AA$ and $6\AA$ in terms of FWHM, respectively.
The data reduction was carried out with the IRAF software\footnote{IRAF is distributed by the National Optical Astronomy Observatories operated by the Association of Universities for Research in Astronomy, Inc. under cooperative agreement with the National Science Foundation.} following standard procedures. The spectra were bias-subtracted and corrected with dome flat-field frames. Cosmic rays were removed interactively from all of the images. Arc-lamp (CuHeNeAr) exposures were used for the wavelength calibration.
Note that, due to the bad sky conditions during the night of March 2015, the spectrophotometric standard was not observed, and therefore the spectrum was not calibrated in flux. %The spectrophotometric standard was observed only for the night of May 2016.

   In the nuclear region, the spectrum (Figure \ref{optical} - right-bottom panel) shows 11 emission lines from which we could derive a redshift $z_{opt}=0.0604\pm 0.0001$, confirming the value of \cite{deGrijp_1992}. The emission line ratios and widths confirm the nature of the core as NLSy1 type AGN.
The spectrum of the northern region (Figure \ref{optical} - right-top panel) is more complex, with emission and absorption lines. The absorption lines present a redshift zero and are typical of a late-type, foreground star that is visible in the optical image of the galaxy (see Figure \ref{optical} - left panel). On the contrary, the emission lines are redshifted to a velocity $+210\: km\,s^{-1}$, relative to the nuclear region ($z=0.0611\pm 0.0002$). The lines are narrow and typical of H\rmnum{2} regions. The uncertainties on the redshifts are the standard deviation.

\section{Analysis of the molecular gas content}
\label{sec:Gas_content}

   \subsection{CO distribution}

\begin{figure*}
  \centering
  \includegraphics[height=4.7cm,trim=35 7 120 0,clip=true]{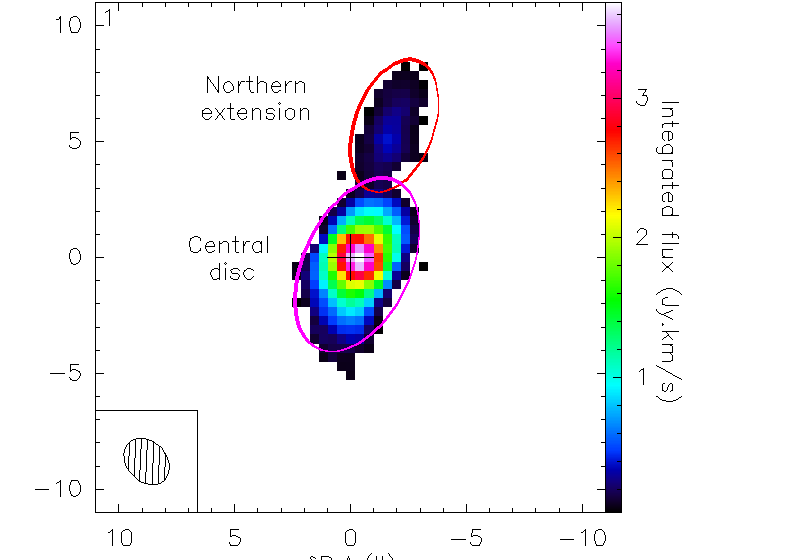}
  \hspace{1mm}
  \includegraphics[height=4.7cm,trim=35 7 80 0,clip=true]{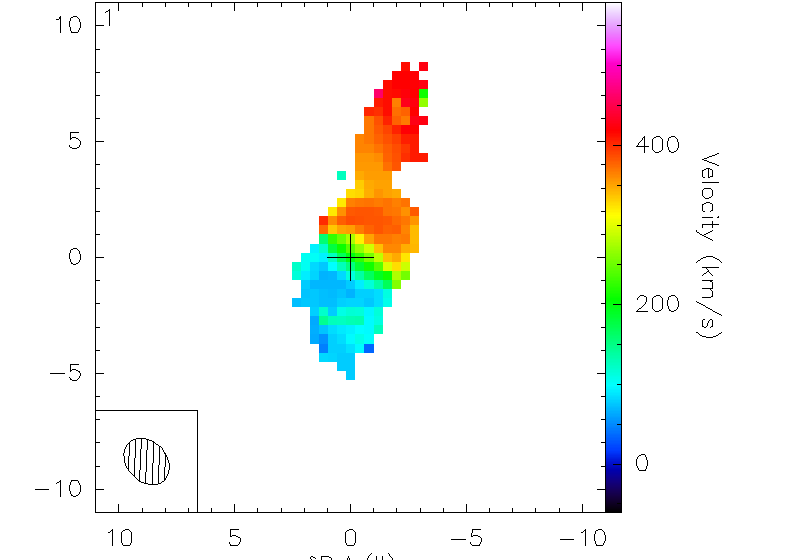}
  \hspace{1mm}
  \includegraphics[height=4.7cm,trim=35 7 80 0,clip=true]{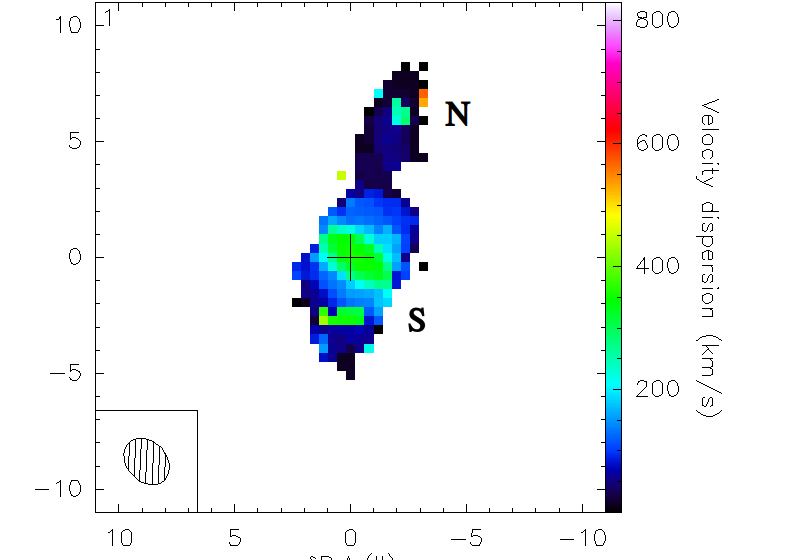}
  \caption{\label{moments} Moment maps of the CO(1-0) emission from NOEMA, as derived from the GILDAS task {\tt moments} using a threshold of $3\sigma$. From \emph{left} to \emph{right}: the integrated intensity in $Jy\,km\,s^{-1}$, the velocity and the velocity dispersion. Most of the molecular gas seems to be distributed within a central apparent disc. The axes show the relative coordinates (in arcsec) with respect to the centre ($\alpha=17^h 03^m 30^s.4$, $delta=+$45:40:47), indicated by the cross. The magenta and red ellipses indicate the central disc and the elongated northern CO emission. We observe two spots at offsets of $[-2.0,5.9]$ (N) and $[0.6,-2.7]$ (S) arcsec where the CO emission has a large velocity dispersion of $250-350\: km\,s^{-1}$.}
\end{figure*}

\begin{figure}
  \centering
  \includegraphics[width=\linewidth,trim=65 35 193 103,clip=true]{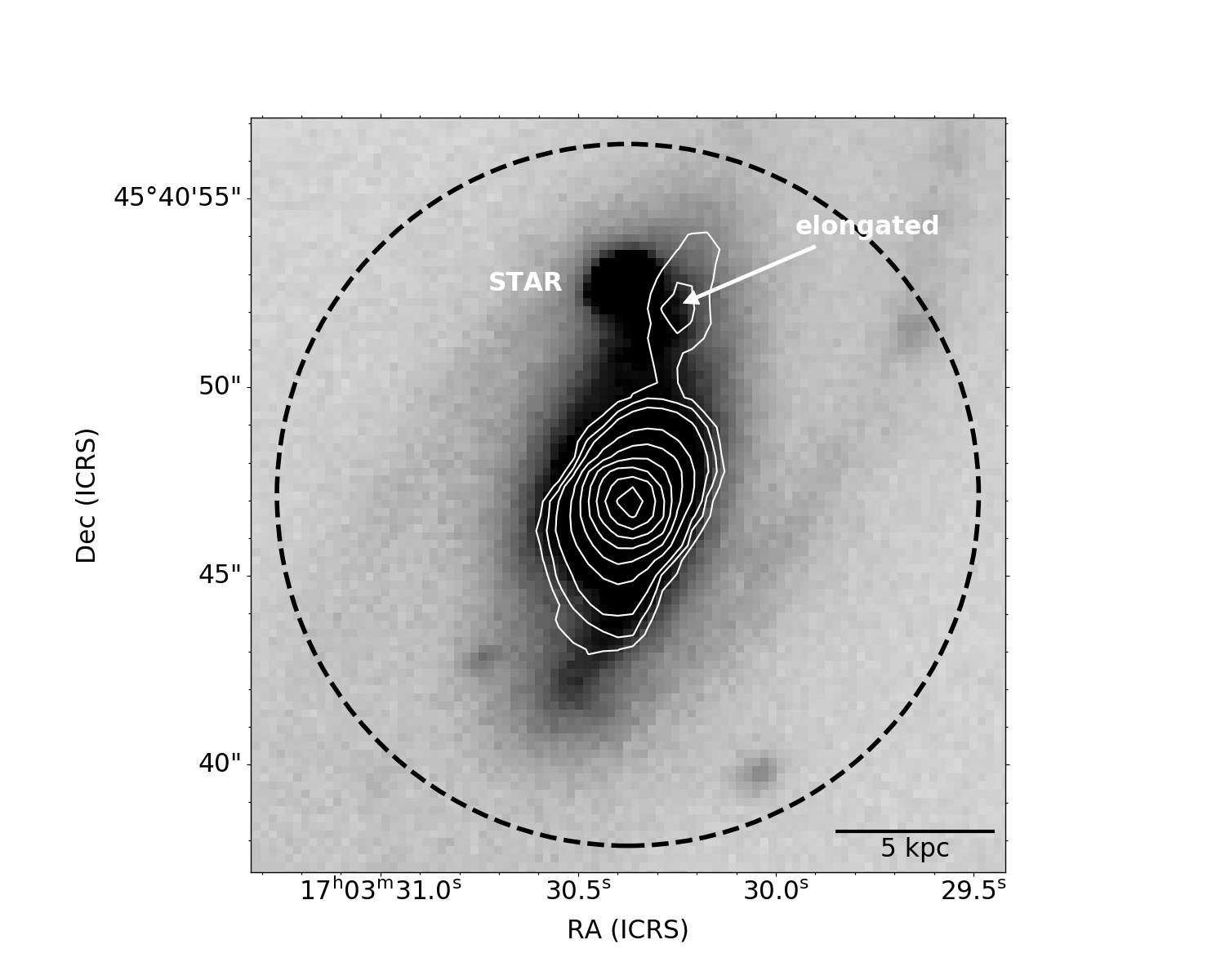}
  \caption{\label{CO-optical} R band image taken by the Nordic Optical Telescope \citep{Olguin_2020} with the contours of the CO emission. The locations of a foreground star and the peak of the elongated CO emission are indicated. The dashed circle shows the beam of the LMT ($18.6''$). We clearly see that the CO emission is associated with the central bar-like structure.}
\end{figure}

   We use the dedicated GILDAS task with a threshold at $3\sigma$ (where $\sigma\sim 0.98\: mJy/beam$) to produce moment maps of the CO emission. After removing the noise-isolated pixels, we obtained the maps presented in Figure \ref{moments}. Most of the CO emission is distributed into a central apparent disc which extends over $\sim 7''$ (magenta ellipse in Figure \ref{moments}), with a similar position angle ($PA\sim -15-20$\degree) that the stellar bar-like structure observed by \cite{Ohta_2007}, suggesting a spatial association (Figure \ref{CO-optical}). A velocity gradient of $\sim 300-350\: km\,s^{-1}$ oriented south-east to north-west is reminiscent of an inclined rotating disc.

   No CO emission is observed in the spiral arms seen in optical (see Figure \ref{CO-optical}). As presented below, NOEMA recovered the same integrated flux as the LMT. The non-detection of CO in the spiral arms could be therefore the result of an intrinsic lack of molecular gas or a too low sensitivity in both the LMT and the NOEMA datasets. Deeper observations would be necessary to answer this question.

   On the opposite, we clearly detected an elongated ($\sim 4''$ long) CO emission centred on an offset of $[-1.2,5.8]$ arcsec north of the nucleus. This northern extension is highlighted by the red ellipse in Figure \ref{moments}. Finally, we identified two small spots, named here N and S, at respective offsets of $[-2.0,5.9]$ and $[0.6,-2.7]$ arcsec which have large velocity dispersions of $250-350\: km\,s^{-1}$. Note that while having similar offsets, the N spot and the northern extension identified in CO emission do not refer to the same region: the N spot being part of the northern extension in projection. In the middle panel of Figure \ref{moments}, the CO emission in the spots N and S is blueshifted and redshifted, respectively, with respect to the surrounding emission.

      \subsubsection{Origin of the high velocity dispersion}

\begin{figure*}
  \centering
  \includegraphics[width=0.48\linewidth,trim=15 25 15 40,clip=true]{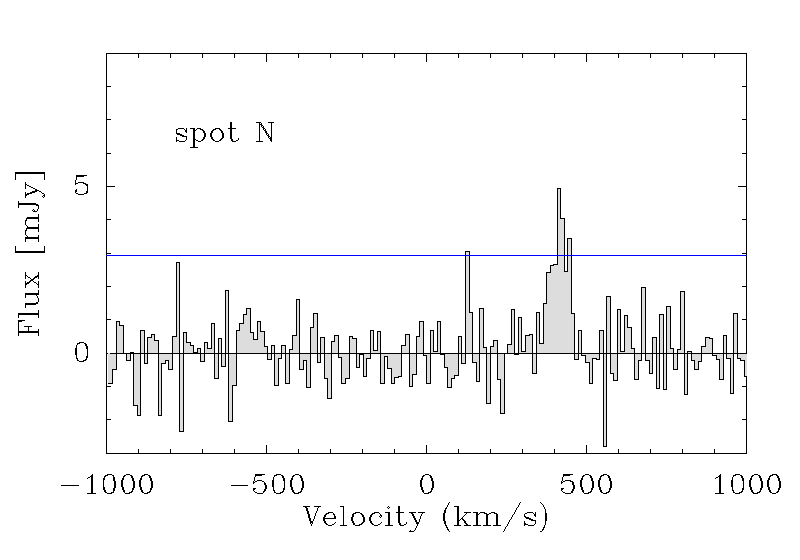}
  \hspace{3mm}
  \includegraphics[width=0.48\linewidth,trim=15 25 15 40,clip=true]{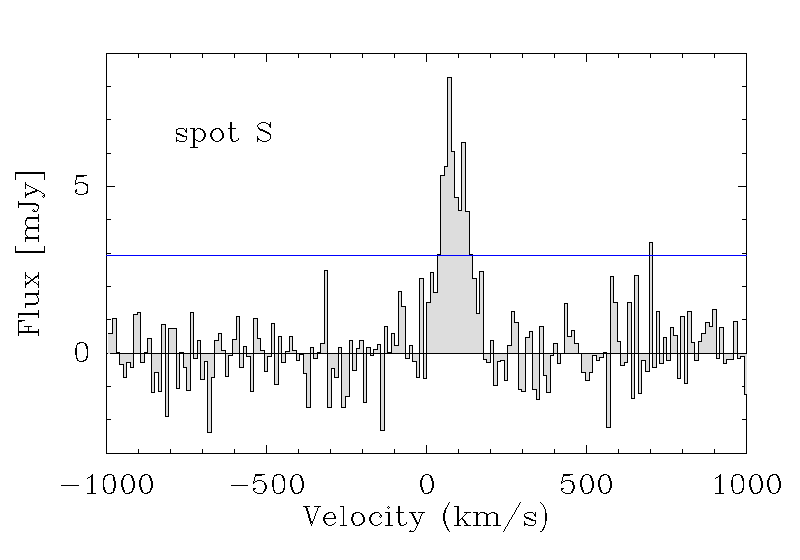}
  \caption{\label{high_vel_disp} CO(1-0) spectrum for the pixel with the highest velocity dispersion in the spot N and S. The horizontal blue line indicates the $3\sigma$ threshold we used to compute the moment maps. The prominent emission line in each spectrum corresponds to the emission from the central disc (\emph{left}) and the northern extension (\emph{right}). We see that isolated peaks lie above the threshold, producing the high velocity dispersion and the shift in velocity with respect to the neighbour pixels.}
\end{figure*}

   Each spot consists in less than 10 pixels for which we extracted the CO(1-0) spectrum. In Figure \ref{high_vel_disp}, we show the spectrum of the pixel with the highest velocity dispersion for each spot. The horizontal blue line shows the $3\sigma$ threshold we used to compute the moment maps. In each spectrum, a strong one-channel, isolated peak (likely noise) lies above the threshold. Therefore, it is taken into account in the calculation of the moments. The high velocity dispersion is thus produced by this isolated peak. Moreover, we note that the peak is blueshifted in the case of the spot N and redshifted for the spot S, with respect to the prominent emission. This also explains the velocity shifts observed in both spots.
We therefore conclude that the spots of high velocity dispersion are not physical and result from noise contamination, with a noisy channel stronger than our $3\sigma$ threshold when producing the moment maps.

   \subsection{Double-peak spectrum}

\begin{figure}
  \centering
  \includegraphics[width=\linewidth,trim=15 25 15 40,clip=true]{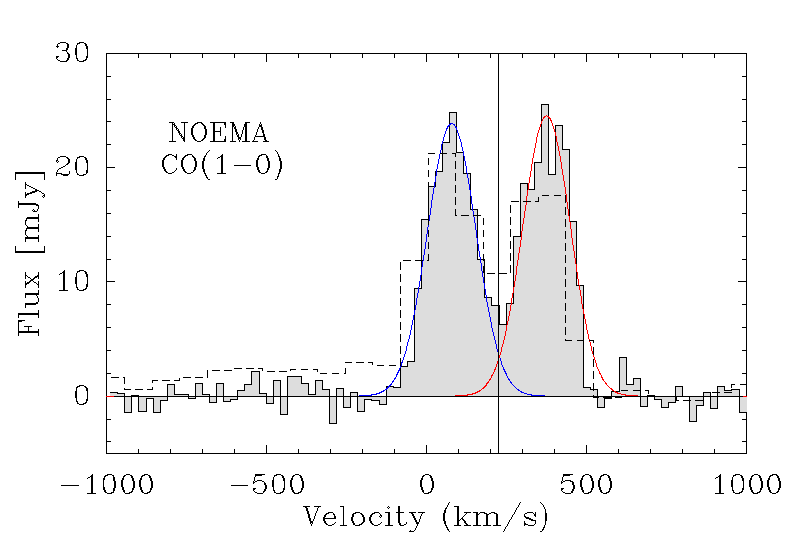}
  \caption{\label{spectra} Spatially-integrated CO(1-0) spectrum from NOEMA, extracted in the region where CO emission is detected (colored pixels in Figure \ref{moments}). For comparison, we overplot the LMT spectrum (dashed line) scaled using the conversion factor $3.5\: Jy/K$. %7 Jy/K in T_A^*
  The zero velocity corresponds to $z=0.0604$ \citep{deGrijp_1992}, while the vertical line indicate the new CO-based redshift ($z=0.0612$).}
\end{figure}

   We now focus on the main galaxy and extracted the spatially-integrated spectrum where CO emission is detected at $3\sigma$ (colored pixels in Figure \ref{moments}). The spectrum of the whole CO(1-0) emission (Figure \ref{spectra}) shows a double-peak profile, similar to the profile previously observed with the LMT \citep{Longinotti_2018a}. For both the LMT and NOEMA spectra, we fitted each velocity components with a Gaussian function to determine its characteristics (Table \ref{table:spec}). The line widths were measured as the full width at half maximum (FWHM), while the fluxes correspond to the integration of the Gaussians. Using the flux density per main beam temperature
ratio for a point source ($S/T_{mb}=3.5\: Jy/K$ for the LMT), we find that the integrated flux of the LMT and NOEMA are consistent within $1\sigma$.

\begin{table}
  \centering
  \small
  \caption{\label{table:spec} Characteristics of the double-peak for both the LMT and NOEMA spectra, corresponding to the characteristics of the Gaussian functions. We have considered a systematic uncertainty of $10\%$ on the absolute flux-calibration.}
  \begin{tabular}{lccccc}
    \hline \hline
    \multicolumn{6}{c}{LMT} \\ \hline
    Line & $T_{mb}$ &   $\Delta v$   &   $v_{peak}$   &     $I_{CO}$      &     $M_{H_2}$      \\
         &   (mK)   & ($km\,s^{-1}$) & ($km\,s^{-1}$) & ($K\,km\,s^{-1}$) & ($10^8\: M_\odot$) \\ \hline
    blue &   6.00   &  $227\pm 22$   &  $62.4\pm 7.9$ &  $1.45\pm 0.26$   &    $5.9\pm 1.1$    \\ %0.11+0.15
    red  &   5.63   &  $182\pm 19$   & $351.6\pm 7.3$ &  $1.09\pm 0.21$   &    $4.5\pm 0.9$    \\ %0.10+0.11
    \hline
    \\ \hline \hline
    \multicolumn{6}{c}{NOEMA} \\ \hline
    Line & $S_{CO}$ &   $\Delta v$   &    $v_{peak}$   &  $S_{CO}\Delta v$  &     $M_{H_2}$      \\
         &  (mJy)   & ($km\,s^{-1}$) &  ($km\,s^{-1}$) & ($Jy\,km\,s^{-1}$) & ($10^8\: M_\odot$) \\ \hline
    blue &   23.9   &  $178\pm 5.0$  &  $78.4\pm 2.1$  &   $4.53\pm 0.56$   &    $6.3\pm 0.8$    \\ %0.11+0.45
    red  &   24.5   &  $175\pm 4.7$  & $374.8\pm 2.0$  &   $4.58\pm 0.56$   &    $6.4\pm 0.8$    \\ %0.10+0.46
    \hline
  \end{tabular}
  \justify {\small
  \textbf{Notes.}
  $S_{CO}$ is the flux density, while $T_{mb}$ refers to the intensity collected in the main beam of the LMT, in units of temperature (i.e. main beam temperature). \\
  For the LMT data, we recalculated the molecular gas mass using the new redshift estimate and $\alpha_{CO}=0.8\: M_\odot\,(K\,km\,s^{-1}\,pc^2)^{-1}$.
  }
\end{table}

   While the peak temperature of the blue component is higher for the LMT spectrum, the height of both components is similar for the NOEMA spectrum. Such difference is likely due to the lower spectral resolution of the LMT observations (about $85\: km\,s^{-1}$) which dilutes the signal into large sampling beams. The low spectral resolution can also explain the difference in the peak velocities (slightly higher with NOEMA) and in the line width of the blue component (larger with the LMT).

\begin{figure}
  \centering
  \includegraphics[width=\linewidth,trim=15 25 15 40,clip=true]{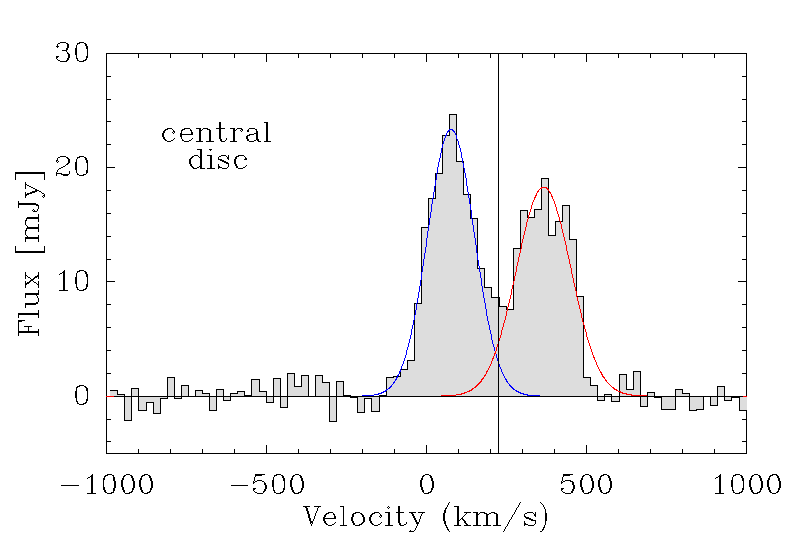}
  \caption{\label{spec_centre} CO(1-0) spectrum for the central disc. The zero velocity corresponds to $z=0.0604$ \citep{deGrijp_1992}, while the vertical line indicate the new CO-based redshift ($z=0.0612$).}
\end{figure}

   We then extracted the spatially-integrated spectrum for the central disc only (magenta ellipse in Figure \ref{moments}). The spectrum also shows a double-peak profile (Figure \ref{spec_centre}), indicating that the molecular gas within the central disc is rotating. However, both CO velocity components have positive peak velocities, while a negative and a positive components are expected in the case of rotation. This indicates that the optical and CO emission do not have the same zero velocity.

\begin{table}
  \centering
  \small
  \caption{\label{table:spec_centre} Characteristics of the CO emission in the central disc. We have considered a systematic uncertainty of $10\%$ on the absolute flux-calibration.}
  \begin{tabular}{lccccc}
    \hline \hline
    Line & $S_{CO}$ &   $\Delta v$   &    $v_{peak}$   &  $S_{CO}\Delta v$  &     $M_{H_2}$      \\
         &  (mJy)   & ($km\,s^{-1}$) &  ($km\,s^{-1}$) & ($Jy\,km\,s^{-1}$) & ($10^8\: M_\odot$) \\ \hline
    blue &   23.4   &  $172\pm 5.1$  &  $75.9\pm 2.2$  &   $4.27\pm 0.54$   &    $5.9\pm 0.7$    \\ %0.11+0.43
    red  &   18.3   &  $201\pm 7.0$  & $365.9\pm 3.1$  &   $3.92\pm 0.51$   &    $5.4\pm 0.7$    \\ %0.12+0.39
    \hline
  \end{tabular}
  \justify {\small
  \textbf{Note.} A conversion factor $\alpha_{CO}=0.8\: M_\odot\,(K\,km\,s^{-1}\,pc^2)^{-1}$ was used for the molecular gas mass.
  }
\end{table}

   The nominal redshifted frequency of our NOEMA observations is 108.705 GHz, based on the optically estimated redshift. Using the CO emission, we find a new systemic velocity (corresponding to the centre of the double-peak) redshifted by $225.6\: km\,s^{-1}$ from the nominal frequency. This corresponds to a redshifted frequency of 108.6233 GHz. We thus get a CO-estimated redshift $z_{CO}=0.0612$, which corresponds to a luminosity distance of 274.3 Mpc. Note that two independent measurements with NOEMA and LMT give the same results on the shape of the molecular emission.
The difference between the optical and CO redshifts is about a factor of 3 larger than the uncertainty on the optical redshift. Therefore, we conclude that this difference is real and might be the result of outflowing gas within the narrow-line region. Future integral field spectroscopy observations will be necessary to investigate in more detail the difference of redshifts.

   \subsection{Molecular gas mass}
   \label{sec:H2_mass}

   In order to estimate the mass of the molecular gas, we used the formulae from \cite{Solomon_1997} for the CO luminosity:
\begin{eqnarray}
  L'_{CO}&=&23.5\, \Omega D_L^2\, I_{CO}\, (1+z)^{-3} \\
  L'_{CO}&=&3.25\times 10^7\, S_{CO}\Delta v\, D_L^2\, \nu_{rest}^{-2}(1+z)^{-1}
\end{eqnarray}
where $\Omega$ is the telescope main beam in arcsec$^2$, $D_L$ is the luminosity distance in Mpc, $I_{CO}$ is the integrated line intensity in $K\,km\,s^{-1}$, $S_{CO}\Delta v$ is the velocity-integrated flux in $Jy\,km\,s^{-1}$, and $\nu_{rest}$ is the rest frequency of CO in GHz. The CO luminosity is given in $K\,km\,s^{-1}\,pc^2$.
Then, applying a CO luminosity-to-H2 mass conversion factor $\alpha_{CO}=0.8\: M_\odot\,(K\,km\,s^{-1}\,pc^2)^{-1}$, typical for ULIRGs \citep{Solomon_2005}, we found a total molecular gas mass $M_{H_2}\sim 1.3\times 10^9\: M_\odot$ within the central disc.
\cite{Lee_2016c} reported an IR luminosity $L_{IR}\sim 3.1\times 10^{11}\: L_\odot$, with AGN contributing to about 44\%. Therefore, after correcting the IR luminosity for the AGN contamination, we derived a star formation rate $SFR\sim 26\: M_\odot\,r^{-1}$ using the formula of \cite{Murphy_2011, Kennicutt_2012}. This gives a molecular depletion time $t_{dep}^{mol}=M_{H_2}/SFR\sim 50\: Myr$, in agreement with the depletion times found by \cite{Combes_2013a} for a sample of 39 ULIRGs. We note that the central disc contains most of the molecular gas with a mass $M_{H_2}\sim 1.1\times 10^9\: M_\odot$ (Table \ref{table:spec_centre}).

%   \subsection{Molecular gas in the northern extension}

\begin{figure}
  \centering
  \includegraphics[width=\linewidth,trim=15 25 15 40,clip=true]{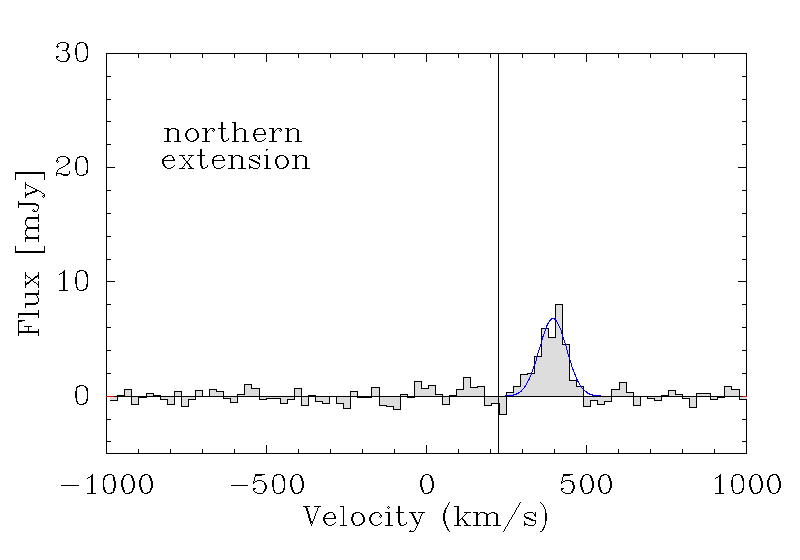}
  \caption{\label{spec_north} CO(1-0) spectrum of the northern extension. The zero velocity corresponds to $z=0.0604$ \citep{deGrijp_1992}, while the vertical line indicate the new CO-based redshift ($z=0.0612$).}
\end{figure}

   We extracted the CO spectrum from the northern extension (Figure \ref{spec_north}). The CO emission shows one single line at a peak velocity of about $395\: km.s^{-1}$ (see Table \ref{table:spec_north}). Therefore, the northern extension is moving away from the central disc at $170\: km.s^{-1}$. The integrated flux is $S_{CO}\Delta v\sim 0.73\pm 0.11\: Jy\,km\,s^{-1}$, which corresponds to a CO luminosity $L'_{CO}=(1.3\pm 0.2)\times 10^8\: K\,km\,s^{-1}$. Assuming the same CO-to-H$_2$ conversion factor as in the main galaxy, we find a molecular gas mass $M_{H_2}=(1.0\pm 0.2)\times 10^8\: M_\odot$.

\begin{table}
  \centering
  \small
  \caption{\label{table:spec_north} Characteristics of the CO emission in the northern extension. We have considered a systematic uncertainty of $10\%$ on the absolute flux-calibration.}
  \begin{tabular}{lcccc}
    \hline \hline
      $S_{CO}$ &   $\Delta v$   &   $v_{peak}$   &  $S_{CO}\Delta v$  &     $M_{H_2}$      \\
        (mJy)  & ($km\,s^{-1}$) & ($km\,s^{-1}$) & ($Jy\,km\,s^{-1}$) & ($10^8\: M_\odot$) \\ \hline
         6.8   &  $101\pm 7.2$  & $394.8\pm 3.1$ &   $0.73\pm 0.11$   &   $1.0\pm 0.15$   \\ %0.04+0.07
    \hline
  \end{tabular}
  \justify {\small
  \textbf{Note.} A conversion factor $\alpha_{CO}=0.8\: M_\odot\,(K\,km\,s^{-1}\,pc^2)^{-1}$ was used for the molecular gas mass.
  }
\end{table}

   \subsection{A molecular outflow in the north}

\begin{figure*}
  \centering
  \includegraphics[height=6.4cm,trim=65 35 190 103,clip=true]{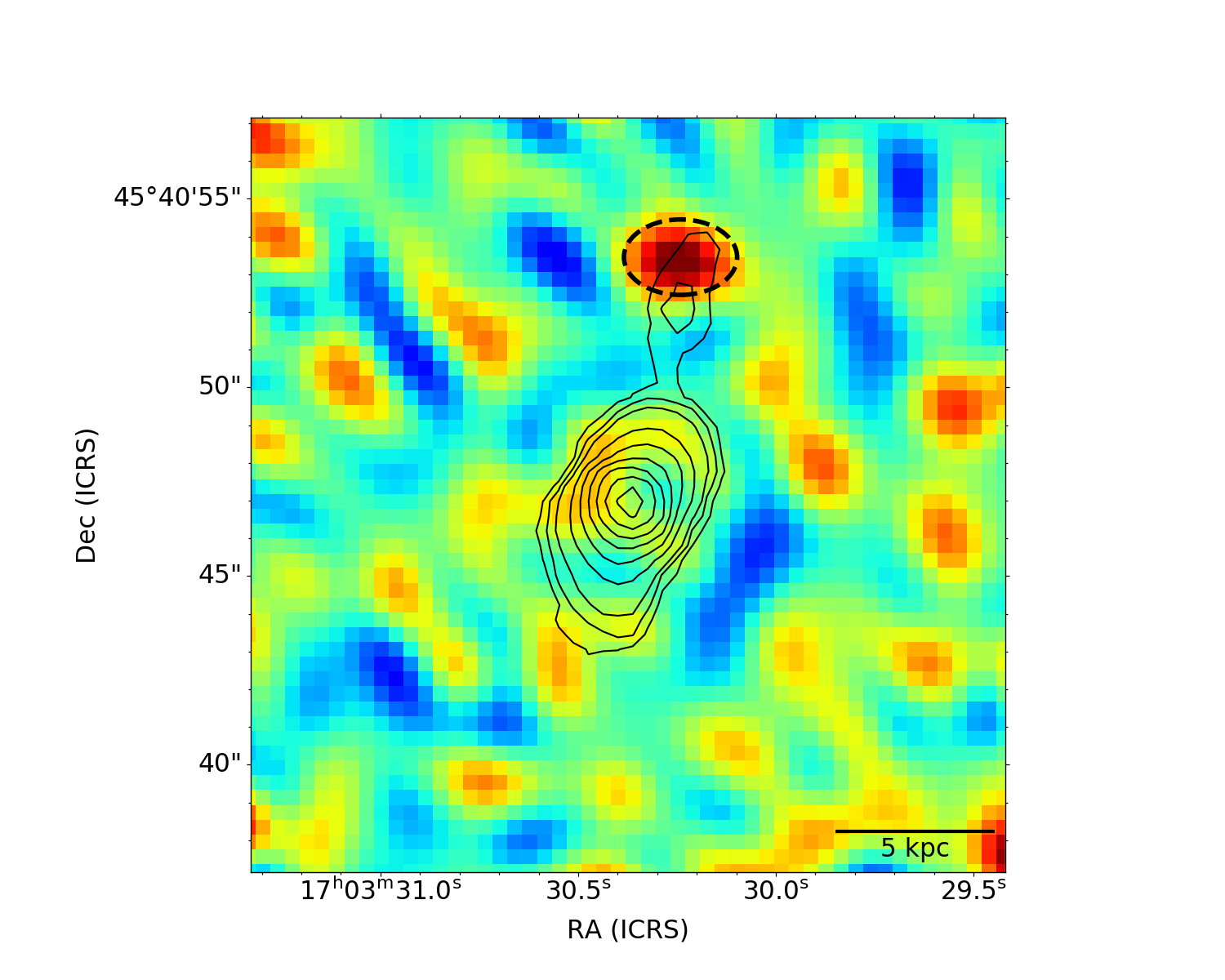}
  \hspace{3mm}
  \includegraphics[height=6.4cm,trim=15 25 15 40,clip=true]{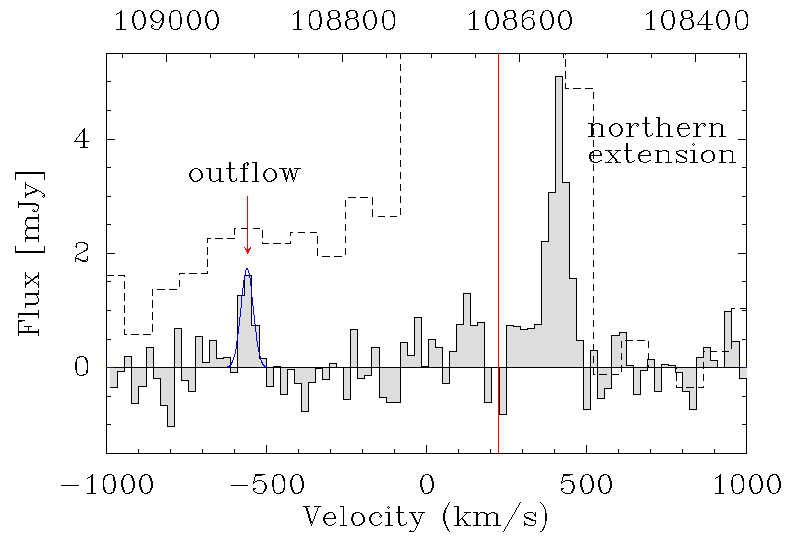}
  \caption{\label{outflow} \emph{Left:} Velocity integrated CO(1-0) emission obtained in the velocity range $-600<v<-500\: km\,s^{-1}$. The black contours show the systemic CO(1-0) emission from Figure \ref{moments}. The black dashed ellipse indicates the position of the bright spot, at an offset of $[-2.0,6.3]$.
  \emph{Right:} CO(1-0) spectrum of the bright spot. The velocity resolution has been decreased to $\sim 22\: km\,s^{-1}$, to reach a rms of 0.47 mJyam. The LMT spectrum is overplotted with the dashed line.
  The zero velocity corresponds to $z=0.0604$ \citep{deGrijp_1992}, while the vertical red line indicate the new CO-based redshift ($z=0.0612$). The emission line at $\sim 400\: km\,s^{-1}$ corresponds to the northern extension which overlaps in projection with the outflow.}
\end{figure*}

   In the left panel of Figure \ref{high_vel_disp}, there is an hint of an extra component in the velocity range $-600<v<-500\: km\,s^{-1}$. We thus investigated this velocity range and found a bright spot of radius $\sim 1\: kpc$ at an offset of $[-2.0,6.3]$, in projection on the northern extension (Figure \ref{outflow}). We extracted the integrated spectrum over the spot. An additional line is detected at $3.5\sigma$ at a velocity of $-561\pm 6\: km\,s^{-1}$ (Figure \ref{outflow}). After redshift correction, the velocity with respect to the main galaxy is $v\sim -787\: km\,s^{-1}$, similar to the velocity of the massive molecular outflow reported in the LMT data by \cite{Longinotti_2018a}

   The integrated flux is $S_{CO}\Delta v=83.6\pm 21.1\: mJy\,km\,s^{-1}$ (Table \ref{table:spec_outflow}), which gives a CO luminosity $L'_{CO}=(1.4\pm 0.4)\times 10^7\: K\,km\,s^{-1}\,pc^2$. Note that the uncertainty is mainly due to the uncertainty on the line width. Estimating the molecular gas mass of the outflow is not straightforward due to the CO-to-H$_2$ conversion factor $\alpha_{CO}$. So far, there is no data set which can be used to evaluate precisely $\alpha_{CO}$ in the outflow of IRAS17. Therefore, we estimate the molecular gas mass using two values of $\alpha_{CO}$: 0.5 (ie. 1/10 the Galactic value \citealt{Feruglio_2010,Longinotti_2018a}) and 0.8 (ie. the typical value for ULIRGS \citealt{Cicone_2012,Cicone_2014,Veilleux_2017}). We therefore derive a molecular gas mass within the outflow of $M_{H_2}=(0.7-1.2)\times 10^7\: M_\odot$ (Table \ref{table:spec_outflow}).
Since the ongoing study of the massive molecular outflow of IRAS17 as seen with NOEMA will be presented in a companion paper (Longinotti et al., in prep.), more details on the properties of this additional outflow will be included and discussed in the context of the outflowing molecular gas within the system.

%%% alpha=0.5 -> M=(7.2$\pm$3.6)$\times$10$^6$ M$_\odot$
%%% alpha=0.8 -> M=(1.2$\pm$0.6)$\times$10$^7$ M$_\odot$

\begin{table}
  \centering
  \small
  \caption{\label{table:spec_outflow} Results of the Gaussian fit for the CO emission within the bright spot at an offset of $[-2.0,6.3]$. Note that the uncertainty in the integrated flux is mainly due to the uncertainty on the line width.}
  \begin{tabular}{lcccc}
    \hline \hline
    $S_{CO}$ &   $\Delta v$   &   $v_{peak}$    &  $S_{CO}\Delta v$   &     $M_{H_2}$      \\
     (mJy)   & ($km\,s^{-1}$) & ($km\,s^{-1}$)  & ($mJy\,km\,s^{-1}$) & ($10^7\: M_\odot$) \\ \hline
      1.74   &   $45\pm 12$   & $-561.2\pm 6.0$ &   $83.6\pm 21.1$    &     $0.7-1.2$      \\ \hline
  \end{tabular}
  \justify {\small
  \textbf{Note.} To calculate the molecular gas mass, we used two value for the conversion factor $\alpha_{CO}$: $0.5\: M_\odot\,(K\,km\,s^{-1}\,pc^2)^{-1}$ (1/10 of the Galactic value) and $0.8\: M_\odot\,(K\,km\,s^{-1}\,pc^2)^{-1}$ (typical of ULIRG).
  }
\end{table}

\section{Analysis of the CO kinematics}
\label{sec:CO_kin}

\begin{figure*}
  \centering
  \includegraphics[height=3.8cm,trim=5 432 5 35,clip=true]{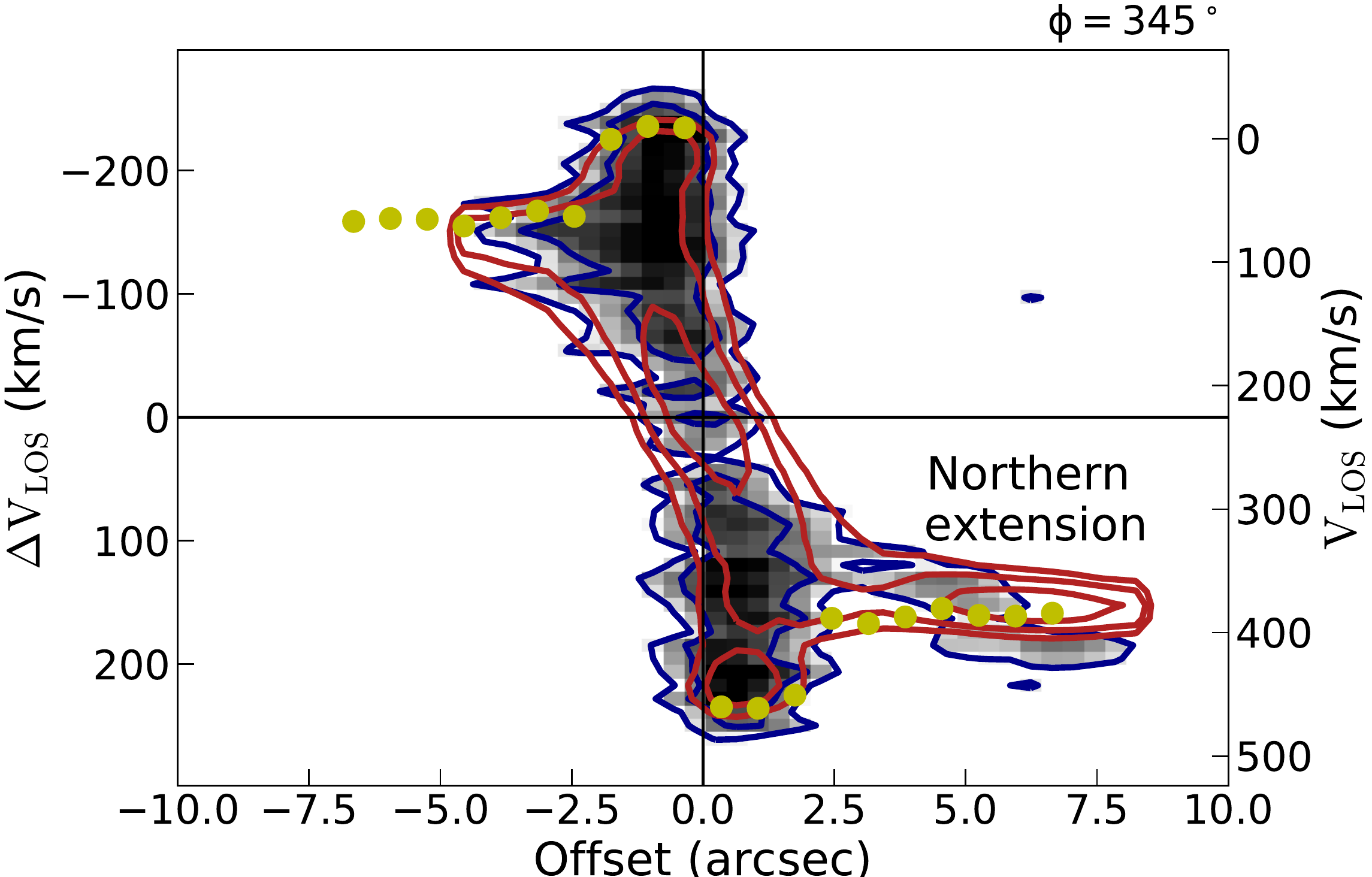}
  \hspace{1mm}
  \includegraphics[height=3.8cm,trim=30 432 30 35,clip=true]{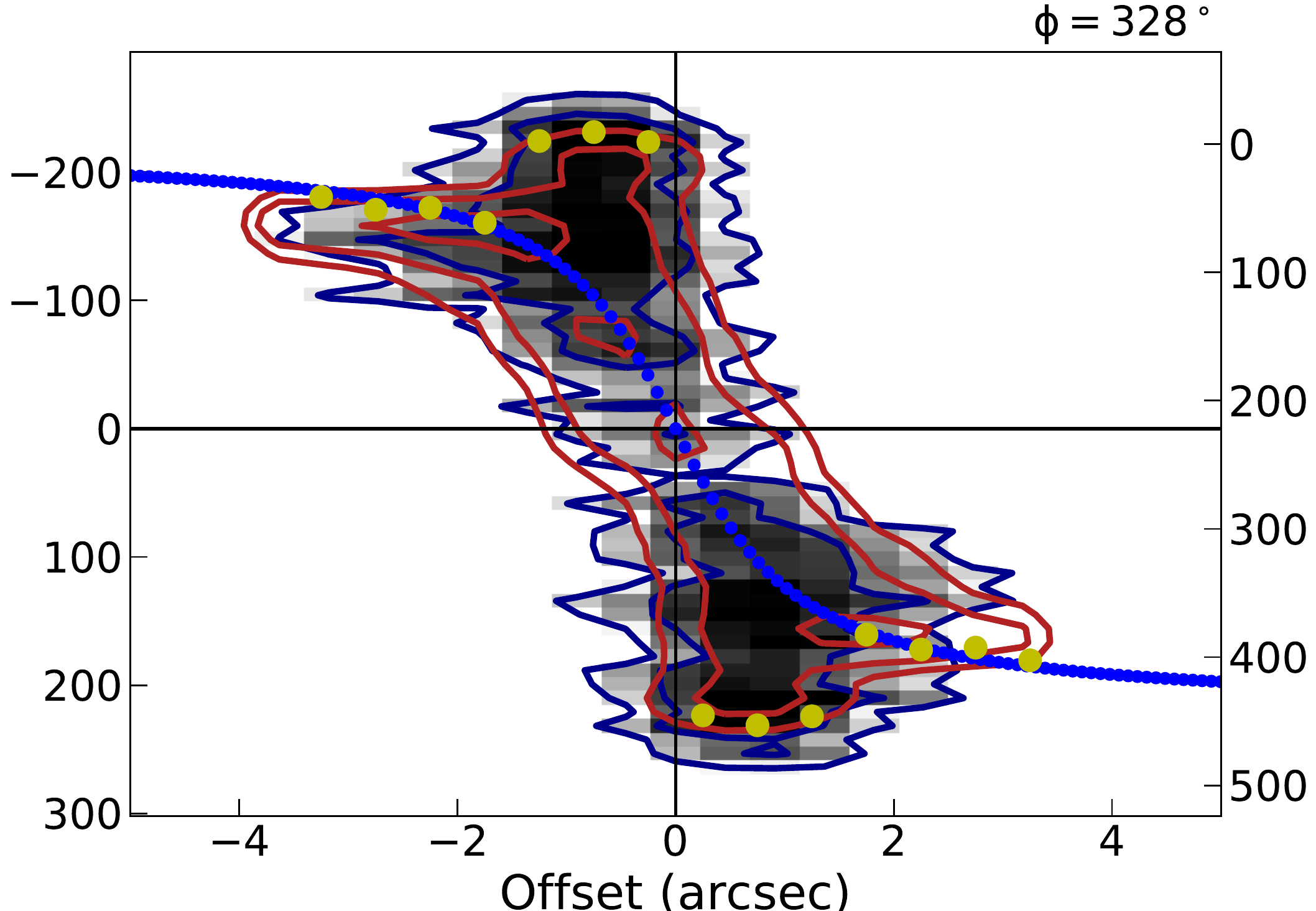}
  \hspace{1mm}
  \includegraphics[height=3.8cm,trim=30 432 30 35,clip=true]{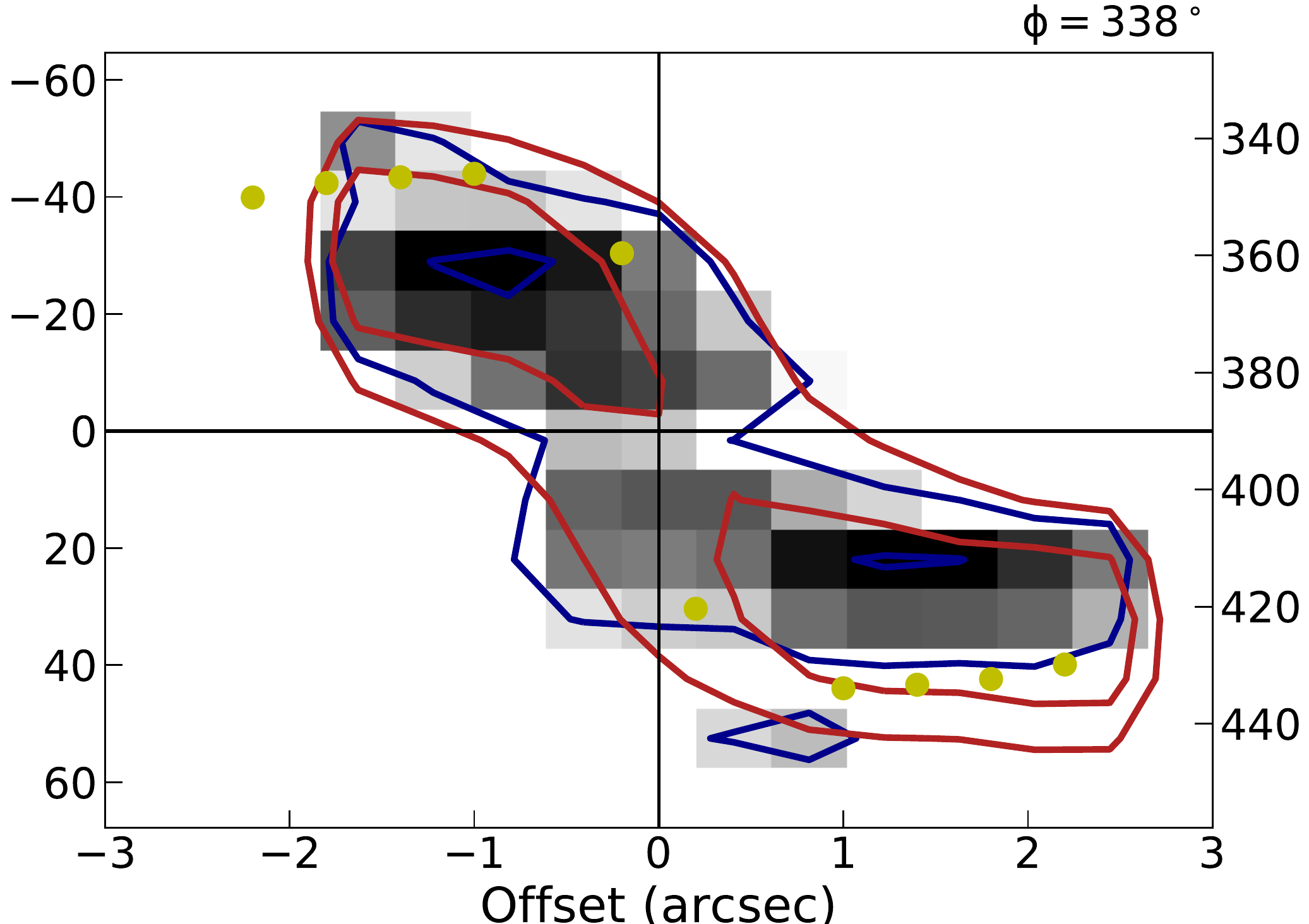}
  \caption{\label{PV_diagrams} Position-velocity diagrams of the CO(1-0) emission for the whole emission (\emph{left}), the central disc (\emph{middle}) and the northern extension (\emph{right}). The blue contours of the observed PV diagram are ($1\sigma$, $2\sigma$, $4\sigma$), while the red ones are the best fit from $^{3D}$BAROLO \citep{diTeodoro_2015}. The y-axes show the same quantities in the three diagrams.
  The yellow dots define the rotation velocity curve. For the central disc, the blue dotted line is the empirical parametrisation by Courteau's function (equation \ref{eq:Courteau}).}
\end{figure*}

   In this section, we will focus on the molecular gas kinematics. We started by producing the position-velocity (PV) diagram  along the major axis ($PA=-15$\degree) of the whole CO(1-0) emission using the "3D-Based Analysis of Rotating Objects from Line Observations" software ($^{3D}$BAROLO; \citealt{diTeodoro_2015}). The northern extension is clearly identified: while the average velocity is consistent with the rotation curve, most of the emission in this region does not follow the rotation (Figure \ref{PV_diagrams} - left panel). We therefore conduct a separate analysis for the central disc (hereafter the main galaxy) and the northern extension.

   \subsection{The main galaxy: a disc or a ring?}

   We used $^{3D}$BAROLO to compute and characterise the PV diagram of the main galaxy along the major axis ($PA=-32$\degree). We run the code using a fixed central position at the brightest pixel in the moment 0 map and a fixed position angle. We enable the inclination to vary in between $65$\degree$\pm 5$\degree.
The PV diagram shows the presence of two components (Figure \ref{PV_diagrams} - middle panel): (1) a "flat" rotation curve reaching $\sim 250\: km\,s^{-1}$ in a few hundred parsecs, and (2) a lower velocity component reaching $\sim 200\: km\,s^{-1}$ at a radius of several kiloparsecs. Such PV diagram is typical of a barred galaxy, where the bar produces higher velocities in the inner region \citep{Binney_1991,Garcia_1995}.
The outer yellow dots in Figure \ref{PV_diagrams} likely correspond to the rotation of the molecular gas. We thus parametrise the rotation curve by fitting these points with the function:
\begin{eqnarray}
  v_c(r) &= v_0\frac{2}{\pi}arctan(r/r_t) \label{eq:Courteau}
\end{eqnarray}
where \{$v_0$, $r_t$\} are free parameters \citep{Courteau_1997,Lang_2020}.

   We then used a simple analytical axisymmetric model which computes the velocity spectrum \citep{Wiklind_1997,SalomeQ_2015a}. Given a radial rotation velocity profile $v_{rad}(r)$ and a gas distribution $n(r)$, the model calculates the velocity spectrum:
\begin{equation} \label{eq:dndv}
  \frac{dN}{dv}(v)=\int \frac{n(r)\, rdr}{v_{rad}(r) \cbra{1-\cbra{\frac{v}{v_{rad}(r)}}^2}^{1/2}}
\end{equation}
Here, we use the rotation curve derived from equation \ref{eq:Courteau}. The gas distribution is described by a Toomre disc of order 2 $n(r)=n_0\cbra{1+\frac{r^2}{d^2}}^{-5/2}$ \citep{Toomre_1964} or a ring, which is the difference between two Toomre discs of caracteristic radii $d_1$ and $d_2$.
To take into account the gas velocity dispersion, we convolve the computed spectrum with a Gaussian of $\sigma^2=\sigma_{obs}^2+\sigma_{ISM}^2$. The observed resolution is $\sigma_{obs}\sim 11\: km\,s^{-1}$ (the channel width). Considering the velocity dispersion $\sigma_{v}\sim 8.5\: km\,s^{-1}$ given by $^{3D}$BAROLO, we therefore convolved by a Gaussian with $\sigma\sim 14\: km\,s^{-1}$.

   We ran a grid of models for a disc/ring of $M_{H_2}\sim 1.1\times 10^9\: M_\odot$ (see Figure \ref{spec_centre} and Table \ref{table:spec_centre}), with an outer characteristic radius $d_1$ from a few hundred parsecs to 6 kpc and an inner characteristic radius $d_2$ from 0 to a few tens of parsecs below $d_1$ ($d_2=0$ corresponds to a disc).
We evaluate the best fitting models by looking at the peak velocity of the double peak. The peak velocity is between $140-150\: km\,s^{-1}$ (Table \ref{table:spec_centre} with correction of the velocity shift). However, to be conservative, we use the central 5\% which correspond to $\sqrt{2}\, erf^{-1}(0.05)\sim 0.063\sigma$. Therefore, we consider the models for which the peak velocity is $145\pm 10\: km\,s^{-1}$.

\begin{figure*}
  \centering
  \includegraphics[page=5,height=7.2cm,trim=45 0 115 30,clip=true]{Model_Courteau_vpeak.pdf}
  \hspace{3mm}
  \includegraphics[height=7.2cm,trim=15 0 40 30,clip=true]{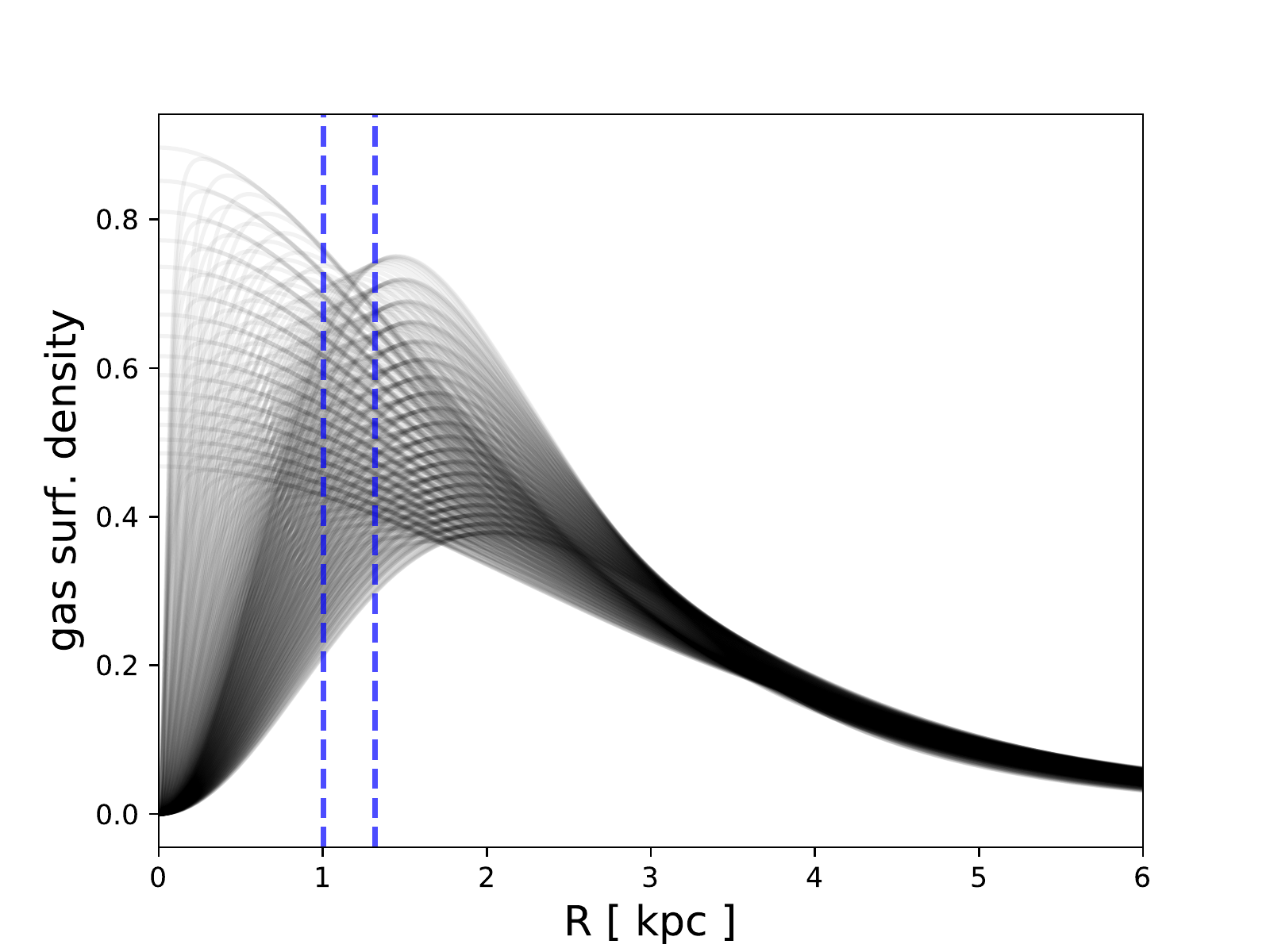}
  \caption{\label{model_results} \emph{Left:} Parameter space of the model grid using the parametrisation of the rotation curve in equation \ref{eq:Courteau}. The outer $d_1$ and the inner $d_2$ characteristic radii vary in the range $0-6.1\: kpc$, with $d_1>d_2$. The white lines are isocontours of the peak velocity, indicating the different gas distributions that fit the observations. The red isocontours represent $R_{peak}$ equal to half the major and minor axis of the beam of NOEMA.
  \emph{Right:} Surface density profiles of the various models within the white isocontours. The y-axis units are arbitrary. The vertical blue lines represent half the major and minor axis of the beam of NOEMA.}
\end{figure*}

   We find that the gas distribution can be described by $d_1=2.4-5.3\: kpc$ and $d2\leq 3.2\: kpc$ (Figure \ref{model_results} - left). This corresponds to either a disc or a ring of molecular gas with surface density-weighted average radii $R_{mean}=1.8-3.1\: kpc$ and a maximum surface density reached at a radius $R_{peak}\leq 2.1\: kpc$ (Figure \ref{model_results} - right). However, an extended ring of gas with $R_{peak}\geq 1\: kpc$ is very unlikely as it would be partially resolved in our observations. Therefore, the best fitting model is likely  $d_1=3.8-5.3\: kpc$ and $d2\leq 0.7\: kpc$. In the case of a ring, this can indicate that the AGN activity affects the molecular gas within the centre, by expelling the molecular gas via a wind/outflow, or by photodissociating the molecular gas.

   Observations at higher resolution will be necessary to disentangle between these two possibilities: disc or ring. Note that we cannot exclude that the molecular gas of the main galaxy is distributed neither in a disc nor a ring, as observed in star-forming barred spiral galaxies \citep{Sheth_2002}.

   \subsection{The northern extension: a dwarf companion}
   \label{sec:PV_north}

   In this section, we investigate the nature of the northern extension. As seen in Section \ref{sec:z_opt}, the optical redshift in this region differs from the central disc by $\Delta z=0.0007$ (Figure \ref{optical}), which corresponds to a velocity difference $\Delta v=210\: km\,s^{-1}$. Moreover, the CO emission within the northern extension is redshifted by about $170\: km\,s^{-1}$ with respect to the centre of the main galaxy (Figure \ref{spec_north}). Finally, as shown in Figure \ref{PV_diagrams}, we do not observe a southern counterpart and the CO emission does not follow the global rotation of the galaxy. All this suggests that the northern extension could be dynamically separated from IRAS\,17020+4544.

   We produced the PV diagram of the northern extension with $^{3D}$BAROLO. The PV diagram reveals a clear rotation pattern with an amplitude of $\sim 50\: km\,s^{-1}$ (Figure \ref{PV_diagrams} - right panel). For a rotation velocity of $50\: km\,s^{-1}$, we derive a dynamical mass $M_{dyn}\sim 1.3\times 10^9\: M_\odot$, one order of magnitude larger than the molecular gas. This is suggesting that the rotation is likely driven by a dark matter halo, strenghtening our hypothesis on the presence of a previously unknown dwarf galaxy.

\begin{figure}
  \centering
  \includegraphics[width=\linewidth,trim=23 191 22 211,clip=true]{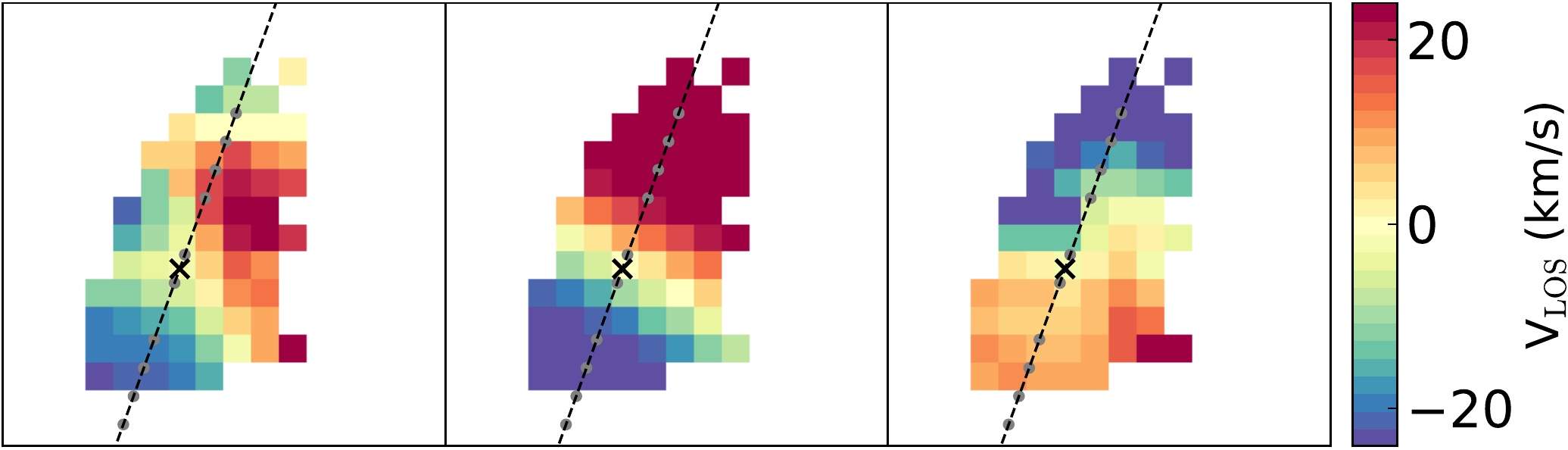}
  \caption{\label{residual_Barolo} Velocity map of the northern extension showing, from \emph{left} to \emph{right}, the observations, the model and the residual. While the PV diagram clearly shows the rotation of the gas, the residual map reveals that the rotation is distorted.}
\end{figure}

\begin{figure*}
  \centering
  \includegraphics[width=\linewidth,trim=50 70 170 50,clip=true]{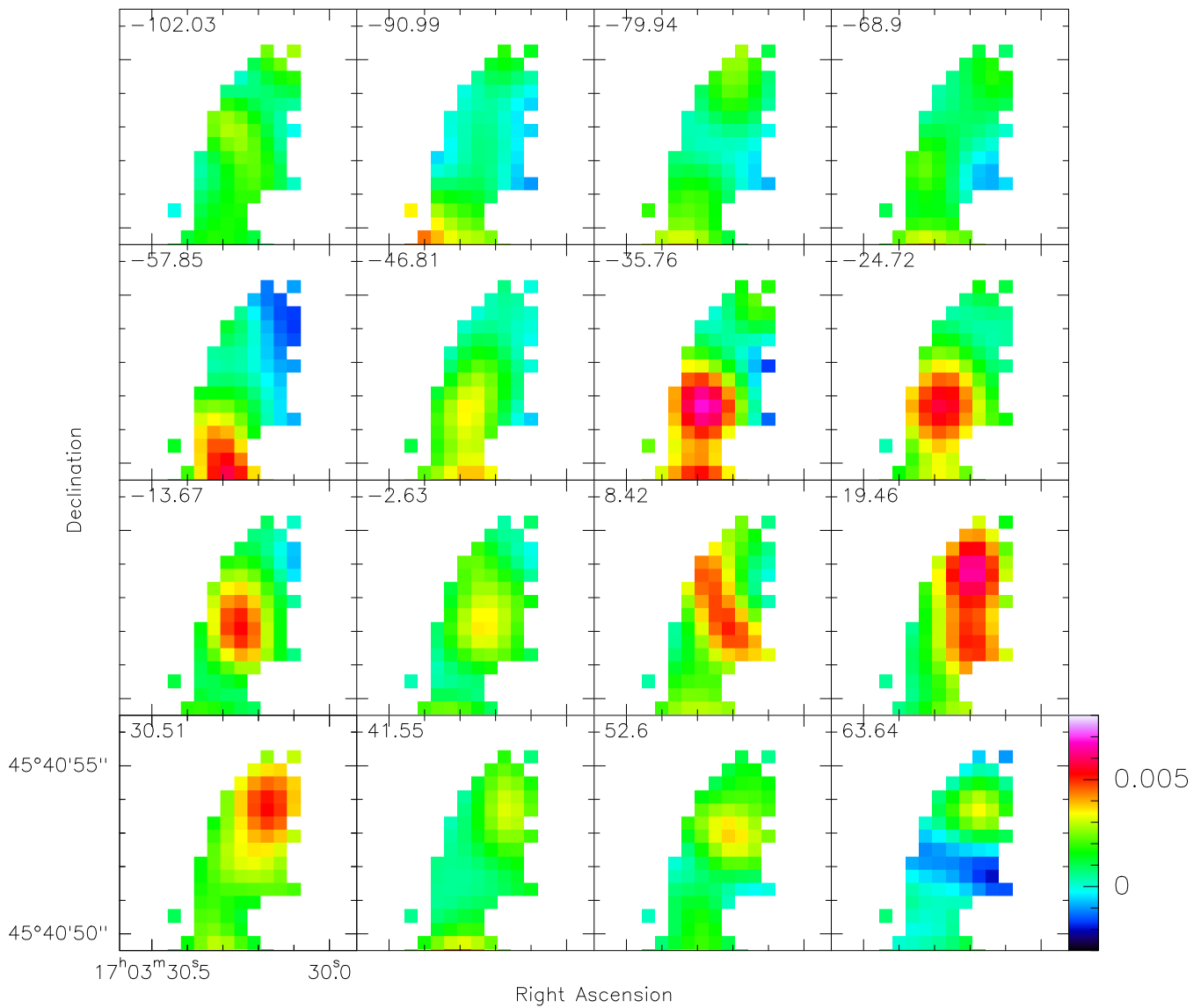}
  \caption{\label{chan_map} Channel map of the CO emission in the companion galaxy. The velocity of the channels, relative to the peak velocity of the CO spectrum $v_{peak}\sim 395\: km\,s^{-1}$, is indicated on the top-left corner. The color bar shows the flux density for each channel, in units of $mJy/beam$. Although the dynamics is governed by rotation, high distortion is evident. At $v=-35.76\: km\,s^{-1}$, there seems to be a "bridge" of material with the main galaxy.}
\end{figure*}

\section{Discussion and conclusion}
\label{sec:conclusion}

   In this article, we have presented new NOEMA observations of the CO(1-0) emission within the nearby NLSy1 IRAS\,17020+4544. These observations allow the spatially resolved properties of the molecular gas content within the galaxy to be characterised for the first time. Most of the CO emission is distributed within an apparent central disc of radius $\sim 4\: kpc$ and of mass $M_{H_2}\sim 1.1\times 10^9\: M_\odot$. Molecular gas was also found up to 8.2 kpc north from the centre. 

   In this northern extension, the molecular gas has a mass of about $10^8\: M_\odot$ (see Section \ref{sec:H2_mass}), much larger than the typical mass of a giant molecular cloud ($10^4-10^6\: M_\odot$, e.g. \citealt{MAMD_2017a}). The PV diagram reveals that the gas dynamics is dominated by rotation (see Section \ref{sec:PV_north}), however this rotation is highly distorted (see Figure \ref{residual_Barolo}). Finally, a "bridge" of material between the northern extension and the central disc is observed (Figure \ref{chan_map}), at a velocity $v=-35.76\: km\,s^{-1}$ relative to the northern extension ($v=133.64\: km\,s^{-1}$ relative to the central disc). All this indicates that the CO emission in the northern extension likely comes from a companion galaxy of diameter $\sim 4.5\: kpc$ which is interacting with IRAS17.
This gives a completely new vision of IRAS17. Before the present discovery, it was difficult to reconcile the presence of multi-phase outflows, as traced by the X-ray UFO and the massive molecular wind, in an apparently unperturbed and perfectly standard spiral galaxy. Instead, it is now conceivable that the interaction plays a major role in triggering the nuclear activity via accretion onto the central super massive black hole.

   Nevertheless, IRAS17 does not show noticeable disturbances in optical images, neither in the dynamics of the CO emission. Therefore, the two galaxies are probably at an early phase of a minor merger (first passing?), which could have driven gas to the central region of the main galaxy, producing a starburst (making IRAS17 a LIRG) and initiating the active phase.
Recently \cite{Olguin_2020} reported deep near-infrared imaging for 29 radio-loud NLSy1 which is strongly indicating that their hosts are preferentially disc galaxies with signs of interaction. This result tells us about the importance of the interaction process in triggering the relativistic jets since a clear majority of the jetted NLSy1 studied so far are found in interacting systems (see also \citealt{Anton_2008, Jarvela_2018,Berton_2019}). However, for IRAS17, \cite{Olguin_2020} did not report any sign of interaction, which supports the idea of an early phase of a minor merger.

   Note that the optical emission lines at the northern position (spectrum A in Figure \ref{optical}) are blueshifted by $\sim 185\: km\,s^{-1}$ with respect to the CO emission. However, the line ratios in this region are typical of ionization by massive stars. The velocity shift thus tends to indicate that the optical emission is unlikely associated with H\rmnum{2} regions. This suggests that the optical emission might not be coming from the companion galaxy, but from the main galaxy. With the present optical observations it is not possible to conclude about this velocity difference, but proprietary data recently collected with the optical integral field spectroscopy instrument MEGARA on the GTC telescope may shed light on this issue (Robleto-Or\'us et al., in prep.).

   The resolution of the NOEMA observations enabled us to study the gas dynamics within the main galaxy. The PV diagram is typical of a barred galaxy. We fitted the rotational part of the PV diagram, which at large galactic radii reaches an asymptotic rotation velocity around $200\: km\,s^{-1}$. The analytical axisymmetric model of \cite{Wiklind_1997} was used to compute the velocity spectrum, given a radial rotation velocity profile and a gas distribution. This simple model reveals that the molecular gas as traced by the CO emission is distributed within a disc or a ring with an inner radius smaller than the resolution of our observations. 

   By applying a new modeling of the morphology of the host galaxy on the NOT observations from \cite{Olguin_2020}, Olgu\'in-Iglesias et al. (in prep.) are finding that their model predicts a disk-like (pseudo-)bulge morphology with presence of a bar. Their results show that that the bar is long (according to Hubble types SB0-SBa), while the bar strength is rather average along all Hubble types. Moreover, the surface brightness matches the morphology of a barred spiral galaxy, namely SBa. This is challenging the distribution of the CO emission in IRAS17, as the molecular gas usually presents a more complex distribution in star-forming barred spiral galaxies \citep{Sheth_2002}.
Higher resolution is now necessary to disentangle between the two scenarios (disc/ring vs. complex distribution) and study the effect of the central AGN on the host galaxy ISM at small scales. The MEGARA data will bring complementary information, which is critical to better understand the host galaxy.

   Finally, we have detected a molecular outflow of $(0.7-1.2)\times 10^7\: M_\odot$ within the north, in projection on the companion galaxy, with a radial velocity in agreement with that of the outflow detected with the LMT by \cite{Longinotti_2018a}. The properties of this outflow and its possible connection with the X-ray UFO will be discussed in a companion paper (Longinotti et al., in prep.) dedicated to the study of the massive molecular outflow as observed by NOEMA.

\section*{Acknowledgements}

We thank the referee for his/her comments. We also thank \'Africa Castillo-Morales for the constructive discussions.

This work is based on observations carried out under project number W17CR with the IRAM NOEMA Interferometer. IRAM is supported by INSU/CNRS (France), MPG (Germany) and IGN (Spain).

Q.S. and A.L.L. acknowledge support from CONACyT research grant CB-2016-01-286316.

YK acknowledges support from UNAM-DGAPA-PAPIIT grant IN106518 and PAPIIT-PASPA.

This work was partially supported by CONACyT research grant 280789.

\section*{Data availability}

The NOEMA data underlying this article will be shared on reasonable request to the corresponding author.

\newcommand{\newblock}{}
\bibliography{Biblio,Biblio_arXiv}
\bibliographystyle{mnras}

\label{lastpage}
\end{document}